%% file: draft-v3.tex
\documentclass[reprint, superscriptaddress,showpacs,nofootinbib,preprintnumbers, aps, prd, longbibliography]{revtex4-2}
\usepackage[utf8]{inputenc}
\usepackage[T1]{fontenc}
\usepackage{quantikz}
\usepackage{multirow}
\usepackage{amsmath, amssymb, makeidx, graphics, amsthm, mathrsfs, color, hyperref, physics, mdframed, appendix, subfiles, titletoc, float, graphicx, enumerate, wrapfig, booktabs}
\input{preamble}

\begin{document}
\preprint{CPTNP-2026-014}
\title{Exponentially improved quantum simulation of scalar QFT}
\author{Qing-Hong Cao}
\email{qinghongcao@pku.edu.cn}
 \affiliation{School of Physics, Peking University, Beijing 100871, China}
 \affiliation{School of Physics, Zhengzhou University, Zhengzhou 450001, China}
 \affiliation{Center for High Energy Physics, Peking University, Beijing 100871, China}

 \author{Ying-Ying Li}
 \email{liyingying@ihep.ac.cn}
 \affiliation{Institute of High Energy Physics, Chinese Academy of Sciences, Beijing 100049, China}

 \author{Xiaohui Liu}
 \email{xiliu@bnu.edu.cn}
 \affiliation{School of Physics and Astronomy, Beijing Normal University, and Key Laboratory of Multiscale Spin Physics (Beijing Normal University), Ministry of Education, Beijing 100875, China}
 \affiliation{Southern Center for Nuclear Science Theory (SCNT), Institute of Modern Physics, Chinese Academy of Science, Huizhou 516000, China}

 \author{Liang-Qi Zhang}
 \email{liangqizhang@pku.edu.cn}
 \affiliation{School of Physics, Peking University, Beijing 100871, China}
 
 \author{Ke Zhao}
 \email{ke-zhao@pku.edu.cn}
 \affiliation{School of Physics, Peking University, Beijing 100871, China}

\date{\today}

\begin{abstract}
Quantum simulations of scalar quantum field theories (QFT) provide important benchmarks for demonstrating quantum advantage. We revisit digitization in the occupation basis, which is typically hindered by unfavorable circuit depth scaling. We present an approach that achieves exponential reductions in circuit depth and significantly mitigates Trotter errors by diagonalizing field operators prior to their decomposition into Pauli strings. Focusing on a scalar QFT in $2+1$ dimensions, we show that this method substantially reduces circuit depth and CNOT gate counts for time evolution. Using the Lorentzian energy--energy correlator as a benchmark observable, we find parameter regimes in which occupation-basis digitization converges more rapidly with respect to local truncation than the amplitude-basis approach of Jordan, Lee, and Preskill. These results provide both algorithmic advances and phenomenological benchmarks for studies of light-ray observables on near-term quantum devices.
\end{abstract}

\maketitle
\section{Introduction.} 
Monte Carlo methods in Euclidean lattice field theory have been highly successful for computing equilibrium observables. However, real-time observables and out-of-equilibrium dynamics remain intrinsically challenging, with similar limitations arising in sign-problem-afflicted regimes. These challenges strongly motivate the development of quantum simulation approaches to quantum field theory \cite{Banuls:2019lgt, Bauer:2022hpo, Funcke:2023lftreview, DiMeglio:2023nsa, Fang:2024ple}.

A concrete quantum simulation framework was established by Jordan, Lee, and Preskill (JLP)~\cite{Jordan:2011ci,Jordan:2012xnu} for real scalar field theory, one of the simplest interacting quantum field theories. In this approach, the real scalar field $\phi(\bm{x})$ is discretized on a spatial lattice, and the local degrees of freedom are digitized in the amplitude basis (AB), defined by $\hat{\phi}(\bm{x})\ket{\phi(\bm{x})} = \phi(\bm{x})\ket{\phi(\bm{x})}$. Digitization is achieved by truncating the local field value $\phi(\bm x)$ to a finite set $\{-\phi_{\rm max}, \phi_{\rm max}\}$ with spacing $\delta_\phi$, thereby rendering the local Hilbert space finite-dimensional. This truncation maps each lattice site to $n_q = \lceil\log_2\!\left(2\phi_{\rm max}/\delta_\phi + 1\right)\rceil$ qubits. Time evolution can be implemented efficiently by taking advantage of the polynomial complexity of the quantum Fourier transformation~\cite{Nielsen:2000}. The AB digitization also exhibits exponential convergence with respect to the truncation size $n_q$ once the Nyquist--Shannon sampling criterion is satisfied~\cite{Somma:2015bcw,Macridin:2018gdw,Macridin:2018oli,Klco:2018zqz}, ensuring rapidly suppressed digitization errors for smooth field configurations.

Despite these advantages, initial state preparation and measurement---such as extracting occupation numbers in momentum modes---require substantial resources even in the free theory~\cite{Jordan:2011ci,Jordan:2012xnu}. This motivates alternative digitization methods. In particular, digitization in the occupation basis (OB) provides more direct algorithms for state preparation and the measurement of occupation numbers~\cite{Klco:2018zqz,YeterAydeniz:2018sqft}, making it attractive for practical implementations in the Noisy Intermediate-Scale Quantum (NISQ) era. However, this simplification comes at a cost. Unlike the AB, the convergence properties for digitization in the OB with respect to local truncation size remain to be systematically understood. Moreover, digitization in the OB leads to non-local interaction terms, exhibits exponential scaling of quantum resources with respect to the local system size, as well as substantial Trotter errors when trotterization is used for implementing the time evolution operator. 

In this work, we propose improved quantum algorithms to partially address the latter challenge. In standard approaches, the interaction Hamiltonian is decomposed into Pauli strings, and its corresponding time-evolution operator is implemented via Trotterization, which leads to large circuit depth and incurs Trotter errors. Our central idea is to diagonalize the truncated field operator prior to performing the Pauli decomposition. This renders the interaction Hamiltonian as a sum of mutually commuting $Z$-type strings, thereby exponentially reducing the number of Pauli terms in the decomposition. Consequently, the corresponding time-evolution operator can be implemented without Trotterization, resulting in a reduced circuit depth while eliminating Trotter errors.

Building on these algorithmic advances, we present a concrete implementation of the interaction time-evolution operator for OB digitizations of $2+1$ scalar QFT and
quantify the associated circuit resources on finite lattices and compare them with those of direct Pauli-string constructions~\cite{Klco:2018zqz} without diagonalization.
As a representative example for calculating Lorentzian observables in QFT, we consider the energy--energy correlator (EEC), which has a long history in collider physics and admits a clean operator formulation in relativistic quantum field theory~\cite{Basham:1978prl,Basham:1978prd,Basham:1979prd,HofmanMaldacena:2008,Kravychuk:2018lightray,Kologlu:2021lightrayOPE,Neill:2022eecreview,Moult:2025energyreview,Lee:2024jnt}.
As a physics benchmark, we evaluate the energy-flow matrix elements entering the EEC on a $2\times2$ lattice, and compare the convergence of the OB digitization with respect to local truncation against the AB baseline of JLP. We also simulate the energy-flow operator on a noiseless quantum simulator and estimate the required quantum resources. Together with convergence studies, these results provide guidance for future investigations of EEC observables on NISQ devices.

This paper is organized as follows. In Sec.~\ref{sec:sft}, we review scalar field theory and its digitization in the OB. In Sec.~\ref{sec:QA}, we discuss resource reduction and Trotter errors, comparing with direct Pauli-string decompositions, and discuss approximate implementations of the time-evolution operator in the asymptotic regime with large truncation size. In Sec.~\ref{sec:eec}, we present the implementation of energy-flow observables and analyze convergence relative to the AB approach of JLP. We then conclude in Sec.~\ref{sec:conclusion}.

\section{Scalar Field Theory.}
\label{sec:sft}
We discretize a real scalar field on an \(N^d\) spatial lattice defined by a
\(d\)-dimensional spatial lattice with lattice spacing \(\delta x\) and
periodic boundary conditions (PBC). The set of lattice sites is
\[
\Lambda = \left\{ \bm{x} = \delta x\,\bm{v}
\;\middle|\;
\bm{v} = (v_1,\dots, v_d), v_i = 0,\dots,N-1\right\},
\]
so that the spatial extent in each direction is
\[
L = N\,\delta x.
\]
The imposition of PBC ensures translational invariance and allows for a
well-defined momentum operator.

The lattice Hamiltonian is given by
\begin{eqnarray}
H &=& \frac{(\delta x)^d}{2} \sum_{\bm{x}\in\Lambda}
\Bigl[
\Pi(\bm{x})^2
- \phi(\bm{x}) \nabla^2 \phi(\bm{x})
+ m_0^2 \phi(\bm{x})^2
\Bigr]
+ H_{\rm int}
\notag\\
&\equiv& H_0 + H_{\rm int},
\label{eq:lattice-Hamiltonian}
\end{eqnarray}
where $\nabla^2$ is the lattice Laplacian operator acting on the field operator and compatible with periodic boundary conditions.

The canonical equal-time commutation relations are
\[
[\phi(\bm{x}), \Pi(\bm{y})]
= \frac{i}{(\delta x)^d}\,\delta_{\bm{x},\bm{y}}.
\]

The discrete momentum set compatible with periodic boundary conditions in the first Brillouin zone is
\[
\tilde{\Lambda}
= \left\{
\bm{p} = -\bm p_{\rm max} + \frac{2\pi}{L}\,\bm{v}
\right\},
\]
with $\bm p_{\rm max} = (\pi/\delta x, ..., \pi/\delta x)$. The non-interacting Hamiltonian $H_0$ can be diagonalized by introducing bosonic ladder operators $a_{\bm p}$ and $a^\dagger_{\bm p}$ corresponding to the creation and annihilation of quanta of the scalar field.

For an $N^d$ lattice, there exist $N^d$ distinct momentum values, each associated with a unique set of creation and annihilation operators. The digitization of the scalar field on the lattice is performed by truncating each momentum modes to \( N_{\rm FS} = 2^{n_q} \) fock states, which can be mapped to \( n_q \) qubits via a binary encoding. Specifically, for each momentum mode $\bm p$, the corresponding occupation state $\ket{\bm p, n}$ is mapped to the computational basis state
\[
\ket{b_{n_q-1}, b_{n_q-2}, \ldots, b_0},
\]
where the occupation number $n$ is given by
\begin{equation}
n = \sum_{j=0}^{n_q-1} 2^j b_j,
\label{eq:numdensity}
\end{equation}
with $b_j \in \{0,1\}$.
Here, \( b_j \in \{0,1\} \) corresponds to the eigenvalue of the projector \( (\mathbb{I}-\sigma_z^{(j)})/2 \) acting on the \( j \)-th qubit. The Hilbert space for the scalar fields is constructed as the direct product over all momentum modes. Thus, the total number of qubits required to encode the scalar degrees of freedom is 
\begin{equation}
N_{\rm dat} = N^d n_q.
\end{equation}
This is distinct from the single-particle encoding \cite{Barata:2020jtq}, as well as the one-hot (unary) encoding of \cite{Hardy:2024ric} in which the number of qubits matches the number of occupation basis states. With such truncation, the operator mapping for the creation and annihilation operators for each momentum mode $\bm p$ in the fock state basis $\ket{{\bm p}, n}$ can be written as
\begin{eqnarray}
a_{\bm p} = \sum_{n=1}^{2^{n_q}-1}\sqrt{n}\,|{\bm p}, n-1\rangle\langle {\bm p}, n|,\notag\\
a^\dagger_{\bm p} = \sum_{n=1}^{2^{n_q}-1}\sqrt{n}\,|{\bm p}, n\rangle\langle {\bm p}, n-1|\,
\end{eqnarray}
with the boundary condition $a^\dagger_{\bm p}\ket{{\bm p}, 2^{n_q}-1} = 0$. The commutation relation becomes
\begin{equation}\label{eq:commutation-relation}
\left[a_{\bm p}, a^\dagger_{\bm p'}\right]
= \delta_{\bm p,\bm p'}
\left(\mathbb{I}
- 2^{n_q}\,|\bm p, 2^{n_q}-1\rangle
\langle \bm p, 2^{n_q}-1|\right).
\end{equation}
The extra projector term originates from truncating the local Hilbert space of each bosonic mode to dimension $N_{\rm FS}$
, which breaks the canonical bosonic algebra at the cutoff and introduces a systematic numerical error. 

The truncated field operators acting on a finite-dimensional Hilbert space can be expressed as
\begin{eqnarray}
    \phi({\bm x}) &=& (\frac{\delta p}{2\pi})^{\frac{d}{2}}\sum_{\bm p\in\tilde\Lambda} \frac{1}{\sqrt{2 \omega_{\mathbf{p}}}} \left( a_{\mathbf{p}} e^{-i \bm p \cdot \bm x} + a^{\dagger}_{\mathbf{p}}e^{i \bm p \cdot \bm x} \right),\notag\\
 \Pi({\bm x}) &=& (\frac{\delta p}{2\pi})^\frac{d}{2}\sum_{\bm p\in\tilde\Lambda} -i\sqrt{\frac{\omega_{\mathbf{p}}}{2} }  \left(a_{\mathbf{p}}e^{-i \bm p \cdot \bm x} - a^{\dagger}_{\mathbf{p}}e^{i \bm p \cdot \bm x}\right),
    \label{eq:phi-pi}
	\end{eqnarray}
with the dispersion relation
\begin{equation}
    \omega_{\bm p}=\sqrt{m_0^2+\frac{4}{\delta x^2}\sum_{i=1}^d \sin^2{\frac{p_i \delta x}{2}}}.
\end{equation}

The free Hamiltonian remains diagonal:
\begin{align}
    &H_0=\frac{1}{2}\sum_{\bm p\in\tilde\Lambda} \omega_{\mathbf{p}} \left( a_{\mathbf{p}} a^{\dagger}_{\mathbf{p}} + a^\dagger_{\mathbf{p}} a_{\mathbf{p}} \right)= \sum_{\bm p\in\tilde\Lambda} \omega_{\mathbf{p}}\nonumber\\& 
    \left[ a^{\dagger}_{\mathbf{p}} a_{\mathbf{p}} + \frac{1}{2} \left(\mathbb{I}-2^{n_q}|\bm p, 2^{n_q}-1\rangle\langle \bm p, 2^{n_q}-1|\right)\right].
\label{eq:hamiltonian}
\end{align}

\section{Quantum algorithms for real-time dynamics.}
\label{sec:QA}
Quantum algorithms for simulating real-time dynamics in quantum field theory typically proceed as follows, as originally formulated by JLP~\cite{Jordan:2011ci,Jordan:2012xnu}. One first prepares the vacuum state or a multiparticle state of the free theory. These states are then adiabatically evolved into the corresponding states of the interacting theory by gradually turning on the interactions. Real-time dynamics, including scattering and showering processes, are subsequently simulated. Finally, the interactions are adiabatically turned off, allowing physical observables to be measured.

One advantage of digitization in the OB, compared to the AB employed by JLP, is the availability of straightforward quantum algorithms for preparing initial states in the free theory and for measuring physically relevant observables such as occupation numbers~\cite{Barata:2020jtq, Hardy:2024ric, Ingoldby:2025bdb}. In particular, the vacuum state $\ket{0}$ is directly mapped to the computational basis state $|000\cdots 0\rangle$. A single-particle wavepacket can be written as
\begin{equation}
\ket{\psi} = \sum_{\bm p} f(\bm p)\, a_{\bm p}^\dagger \, |000\cdots 0\rangle,
\end{equation}
where $f(\bm p)$ is a momentum-space distribution function specifying the shape of the wavepacket, typically chosen to be Gaussian.

The action of $a_{\bm p}^\dagger$ can be realized by applying a Pauli-X operator to the qubit representing the single-occupation state in the register corresponding to momentum $\bm p$. Consequently, the operator $\sum_{\bm p} f(\bm p)\, a_{\bm p}^\dagger$, which is a linear combination of unitaries, can be implemented using the Linear Combination of Unitaries (LCU) method~\cite{Childs:2012gwh}.

Multiparticle wavepackets are prepared analogously, with the key difference being the use of higher powers of the creation operator. For a given momentum mode $\bm p$, the action of $(a_{\bm p}^\dagger)^n$ on the vacuum corresponds to setting the occupation-number register of that mode to the integer $n$. This is implemented by expressing $n$ in binary form and applying Pauli-$X$ gates to the qubits corresponding to bits equal to $1$. Superpositions over momentum modes are constructed, as in the single-particle case, using the LCU method.

To extract physical information from the real-time dynamics, such as momentum-space differential cross sections, one directly measures all qubits in the computational basis.

In contrast to the local formulation in JLP, digitization in OB renders the interaction Hamiltonian $H_{\rm int}$ nonlocal in the qubit Hilbert space. As a result, efficient implementation requires all-to-all qubit connectivity, as available in trapped-ion platforms~\cite{Grzesiak:2019qff} and Rydberg-atom-based quantum processors which can feature programmable long-range interactions and can approximate all-to-all connectivity within a blockade radius or via atom rearrangement~\cite{Browaeys:2020,Ebadi:2021,Bluvstein:2022}. This avoids the overhead associated with decomposing long-range two-qubit operations into nearest-neighbor gates.

While the unitary evolution generated by the free Hamiltonian $H_0$ is diagonal in OB and can therefore be implemented efficiently, the interaction Hamiltonian $H_{\rm int}$ remains the primary computational challenge. In the following, we propose a new method to implement the time-evolution operator associated with $H_{\rm int}$, with the goal of demonstrating reductions in circuit depth and Trotter errors for NISQ-era devices. 

\subsection{Algorithm for the time evolution operator}
To implement the time evolution on a quantum circuit, the evolution operator $e^{-i H t}$ can then be approximated using the Suzuki--Trotter formula \cite{Suzuki:1976be}.
\begin{equation}
    e^{-i H t}=\lim_{N_t\to\infty} [e^{-i H_0 \delta t}e^{-i H_{\rm int} \delta t}]^{N_t}
\end{equation}
with $\delta t = t/N_t$. Since \( H_0 \) is diagonal in OB, the time-evolution operator \( e^{-i H_0 \delta t} \) can be implemented in parallel with one-layer of phase-rotation gates. The evolution operator for \( H_{\text{int}} \) is non-diagonal and is typically implemented by decomposing \( H_{\text{int}} \) directly into Pauli strings. With the unitary operator corresponding to each Pauli string implemented with a quantum circuit, \( H_{\text{int}} \) can be realized with trotterization error from noncommuting pairs of Pauli strings \cite{Klco:2018zqz, Hardy:2024ric}. This approach leads to a circuit depth as well as a trotterization error that quickly becomes prohibitively large. 

Consider specifically, the interaction term in $\lambda\phi(\bm x)^s$ theory with,  
\begin{equation}
    H_{\rm int} = \frac{\lambda}{s!}\delta x^d\sum_{\bm x\in\Lambda}\phi(\bm x)^s.
\end{equation}
Neglecting the non-commutativity of $\phi(\bm x)$ operators at different lattice sites arising from finite truncation---an effect that impacts only the highest-occupation state for each momentum mode $\mathbf{p}$, we have
\begin{equation}
\begin{aligned}
    e^{-i H_{\mathrm{int}} \delta t}&=\mathrm{exp}\big(-i\theta\sum_{\bm x\in\Lambda}\phi(\bm x)^s\big)\\ &= \prod_{\bm x\in\Lambda} \mathrm{exp}\big(-i\theta\phi(\bm x)^s\big),
    \label{eq:evo_int}
\end{aligned}
\end{equation}
where the factors of coupling $\lambda$ and $\delta t$ have been absorbed into an overall parameter $\theta$. A worst-case estimation shows that each local interaction term $\phi(\mathbf{x})^s$ decomposes into
\(
\mathcal{O}\!\left(C_{N^d}^{s}\,4^{s n_q}\right)
\)
Pauli strings. The combinatorial factor $C_{N^d}^{s}$ counts the number of ways to select $s$ momentum modes out of the $N^d$ available modes. For each such choice, the resulting operator acts nontrivially on $s n_q$ qubits, giving rise to maximally $4^{s n_q}$ distinct Pauli-string combinations.
The circuit depth of implementing $\exp\big(-i\theta\phi(\bm x)^s\big)$ exhibits the same scaling, differing by the circuit depth required to implement the unitary operator corresponding to a single Pauli string. As involving operations acting on the same set of qubits, the terms at different $\mathbf{x}$ in \eq{evo_int}, despite commuting, cannot be implemented in parallel. Consequently, the total circuit depth scales as \(N_{\rm ps} \sim \mathcal{O}\!\left(N^d C_{N^d}^{s} 4^{s n_q}\right)\). For a worst-case estimate of the errors in implementing \(e^{-i H_{\mathrm{int}}}\)
from trotterization, we consider the maximal number of nonzero commutators scales as \(\mathcal{O}(N_{\rm ps})\), and maximal of the norm is \(\mathcal{O}\!\left(2^{s n_q}\right)\) for a commutator acting on $s n_q$ qubits. Consequently, the worst-case trotterization error scales as $\mathcal{O}\!\left(N^d C_{N^d}^{s} 2^{3s n_q}\right)$.

To overcome these difficulties, we propose the method to implement $e^{-i H_{\mathrm{int}}\delta t}$ based on diagonalization of the field operator \(\phi(\bm x)\). For the operator $a_{\mathbf{p}} e^{-i \bm p \cdot \bm x} + a^{\dagger}_{\mathbf{p}}e^{i \bm p \cdot \bm x}$, we notice that they can be written as $R^\dagger(\bm p\cdot \bm x)(a_{\mathbf{p}}+a^\dagger_{\mathbf{p}})R(\bm p\cdot \bm x)$, with $R(\bm p\cdot \bm x) = e^{- i \bm p\cdot \bm x a_{\mathbf p}^\dagger a_{\mathbf p}}$. The diagonalization of the field operator $\phi(\bm x)$ is then based on the diagonalization of the matrix
$A_{\mathbf p}= a_{\mathbf p} + a_{\mathbf p}^\dagger$. Denote the diagonalized matrix $\tilde{A}_{\mathbf p}$ corresponding to $A_{\mathbf p}$ as
\begin{equation}
    \tilde{A}_{\mathbf p}= \mathrm{diag}(\lambda_1, \lambda_2, \ldots, \lambda_M),
    \label{eq:Adiag}
\end{equation}
and
\begin{equation}
    A_{\mathbf p}=G_{\mathbf p}\tilde{A}_{\mathbf p}G^\dagger_{\mathbf p}
\end{equation}
where $\lambda_i/\sqrt{2}$ is the $i$-th zero of the Hermite polynomial $H_M(x)$, $G_{\mathbf p}$ is the Hermite transformation acting on the subspace spanned by the Fock states with momentum $\mathbf p$.
The field operator $\phi(\bm x)$ thus admits a transformation of the form $\phi(\bm x) = P(\bm x) \,\tilde{\phi}(\bm x ) P^{\dagger}(\bm x)$ with $\tilde{\phi}(\bm x )$ being a diagonal matrix in the occupation basis $\ket{\bm p, n}$,
\begin{eqnarray}
    \tilde{\phi}({\bm x}) &=& (\frac{\delta p}{2\pi})^{\frac{d}{2}}\sum_{\bm p\in\tilde\Lambda} \frac{\tilde{A}_{\bm p}}{\sqrt{2 \omega_{\mathbf{p}}}}.
    \label{eq:phi-pi}
	\end{eqnarray}
The transformation matrix
\begin{equation}
    P(\bm x)=\bigotimes_{\bm p\in\tilde\Lambda}R(\bm p\cdot \bm x)G_{\mathbf p}, 
\end{equation} 
is a tensor product of matrices acting on subspaces of dimension $n_q$ to implement the transformation from the fock state basis to the diagonal basis of $\tilde{\phi}(\mathbf x)$. 
We therefore obtain
\begin{equation}
\begin{aligned}
    e^{-i H_{\mathrm{int}}\delta t} &= \prod_{\bm x\in\Lambda} P(\bm x)\mathrm{exp}\big(-i\theta\tilde{\phi}(\bm x)^s\big)P^\dagger(\bm x).
\label{eq:time_evo_gate}
\end{aligned}
\end{equation}
Since \(P(\mathbf{x})\) is unitary, it admits a direct implementation as a quantum circuit. 
Moreover, the circuits corresponding to different momentum modes can be executed in parallel. Consequently, the circuit depth required to implement $P(\mathbf{x})$ at a given lattice site $\mathbf{x}$ scales as
\begin{equation}
    D_{P,\mathrm{single}} \sim \mathcal{O}\!\left(4^{n_q}\right).
\end{equation}
Summing over all lattice sites, the total circuit depth contribution from this subroutine scales as $D_P \sim \mathcal{O}\!\left(N^d 4^{n_q}\right)$.

We can extend the discussion to the operator $\Pi(\bm x)$, where the matrix that diagonalizes $\Pi(\bm x)$ to $\tilde{\Pi}(\bm x)$ is denoted $P_{\Pi}(\bm x)$.

Since \(\tilde{\phi}(\mathbf{x})^s\) is diagonal, it can be decomposed into Pauli strings consisting only of the identity and Pauli-Z operators. This property yields twofold benefits. Firstly, as the Pauli strings decomposition of $\tilde{\phi}(\bm x)^s$ involves only identity matrix and Pauli-Z operator which commute with each other, the unitary operator $\exp\big(-i\theta\tilde{\phi}(\bm x)^s\big)$ can be implemented exactly, thus eliminating the substantial Trotter errors from direct Pauli string decomposition of $\phi(\bm x)^s$. Secondly, the circuit depth to realize $\exp\big(-i\theta\tilde{\phi}(\bm x)^s\big)$ scales as \(
\mathcal{O}\!\left(C_{N^d}^{s}\,2^{s n_q}\right)
\)
which lead to the total circuit depth including all sites $\tilde{\phi}(\rm x)$ to be \(\mathcal{O}\!\left(N^dC_{N^d}^s 2^{s n_q}\right)\). Together with a volume-independent circuit-depth for implementing \(P(\mathbf{x})\), the method with diagonalization representing a substantial reduction compared to the direct Pauli decomposition approach.

We note that an approximate implementation of the Hermite transformation \(G_{\mathbf{p}}\) has recently been developed~\cite{Jain:2025toh}, achieving circuit depths that scale polynomially with both \(n_q\) and the target accuracy. Combined with the application of single-qubit phase rotations \(R_z\!\left(2\,\bm{p}\cdot \bm{x}\, 2^j\right)\) on the \(j\)-th qubit to implement \(R(\bm{p}\cdot \bm{x})\), this enables an approximate realization of \(P(\mathbf{x})\) with polynomial complexity. However, an efficient implementation of the unitary operator \(\exp\big(-i\theta \tilde{\phi}(\bm{x})^s\big)\) with polynomial scaling in both system size and target accuracy is still lacking. This limitation is expected to become the primary bottleneck in extending OB digitization methods to large systems.

\subsection{Benchmarks}
\begin{table*}[!htbp]
\centering
\caption{The CNOT circuit depth for implementing $e^{-iH_{\rm int}\delta t} (s = 4)$ across different truncation $n_q$, using the method of direct Pauli string decomposition (``Non-diag'') and the diagonalization-based method (``Diag''), where we have chosen $d=2$, $N=2$. The value outside the parentheses corresponds to the reduced circuit depth obtained after optimization with Qiskit.}
\label{tab:cnot_depth_nq}
\begin{tabular}{c|cccc}
\hline
$n_q$ 
&1& 2 & 3 &4  \\
\hline
Non-diag & (64)56   &  (40128)27696  & (2738424)1665752 & (132112108)60975028  \\
\hline
Diag &  (64)56 & (1656)916 &(29552)15704 & (439952)132660  \\
\hline
\end{tabular}
\end{table*}
Consider the case of $s=4$, we compare the CNOT circuit depth using direct Pauli string decomposition (``Non-diag'') and the diagonalization-based method (``Diag'') to implement $e^{-iH_{\rm int}t}$ across different lattice size $N$ and truncation $n_q$. The value outside the parentheses corresponds to the reduced circuit depth obtained after optimization with Qiskit. For a $2d$ spatial lattice with $N = 2$, \tab{cnot_depth_nq} shows the CNOT circuit depth increase with $n_q$. We observe that using the method with diagonalization, the circuit depth for the one-step time evolution of a $d=2$ dimensional lattice with $n_q=2$, $n_q=3$ and $n_q = 4$ can be reduced by a factor of $\mathcal{O}(30)$, $\mathcal{O}(100)$ and $\mathcal{O}(400)$, respectively, comparing to the direct Pauli string decomposition. 

 Fixing $n_q = 2$, \tab{cnot_depth_N} shows the circuit depth for different $N$. Again, the circuit depth from the diagonalization-based method can be reduced by a factor of $\mathcal{O}(30)$ and $\mathcal{O}(60)$. Furthermore, such improvements result in a lattice $2\times 2$ system with $n_q =2$ a realistic benchmark for near-term quantum devices that target $\mathcal{O}(10^3)$ circuit depth. 
\begin{table}
\centering
\caption{Same as \tab{cnot_depth_nq}, but showing the CNOT circuit depth dependence on the lattice size $N$ (here $d=2, n_q=2)$.}
\label{tab:cnot_depth_N}
\begin{tabular}{c|cc}
\hline
$N$ 
& 2 & 3  \\
\hline
Non-diag          & (40128)27696 &(4666356)2912796  \\
\hline

Diag   & (1656)916 & (152955)46746   \\
\hline
\end{tabular}
\end{table}

We can extend this method to time-evolution operators for interaction terms involving fields on different lattice sites, such as
$\phi(\bm{x})\,\phi(\bm{x}+\delta x\,\bm{i})$, as considered in Ref.~\cite{Klco:2018zqz}.
Such terms can arise from the finite-difference discretization of the spatial Laplacian,
$\phi(\bm{x}) \nabla^{2} \phi(\bm{x})$, which is typically absorbed into the free Hamiltonian $H_0$.
To illustrate the efficiency of the diagonalization-based approach, we instead consider the implementation cost of the time-evolution operator for a representative intersite interaction,
$\phi(\bm{x})\,\phi(\bm{x}+\delta x \bm i)$,
on a $1+1$-dimensional lattice.
As shown in Table~\ref{tab:cnot_depth_dphi}, the CNOT counts (in parenthesis) obtained agree with those reported in Ref.~\cite{Klco:2018zqz}. The value outside the parentheses again corresponds to the circuits after optimization with Qiskit. We observe that the diagonalization-based method reduces the CNOT count by a factor of $\mathcal{O}(10)$, while the circuit depth by a factor of $\mathcal{O}(60)$.
\section{Energy flow and energy-energy correlator}
\label{sec:eec}
A comprehensive assessment of the feasibility of quantum simulations targeting specific processes requires consideration of two key resources: the total number of quantum gates needed to implement the simulation and the number of qubits required in the digitization to achieve a given precision. The Nyquist–Shannon sampling theorem, as introduced in Refs.~\cite{Macridin:2018oli, Macridin:2018gdw} building upon the work of Somma~\cite{Somma:2015bcw}, provides guidance for evaluating the efficacy of scalar-field digitization methods. Digitization errors in the ground-state energy have been explicitly analyzed in both the AB basis as employed by JLP framework and OB basis, where higher precision can be achieved in the OB basis with the same number of qubits for localized field-space wave functions~\cite{Klco:2018zqz}.
In addition, the number of qubits required to simulate scattering processes at a given energy scale has been observed to be smaller in the OB basis~\cite{Ingoldby:2025bdb}.

\begin{table}[htbp]
\centering
\caption{The CNOT count and depth to implement the time evolution operator for an interaction term $\phi(\bm x)\phi(\bm x+\delta x \bm i)$. The CNOT count without Qiskit optimization (in parenthesis) agrees with that in \cite{Klco:2018zqz}.
}
\label{tab:cnot_depth_dphi}
\setlength{\tabcolsep}{2pt}
\begin{tabular}{|cc|cc|}
\hline
\textbf{Basis} & \textbf{$n_q$} & \textbf{CNOT count} &\textbf{CNOT depth} \\
\hline
\multirow{5}{*}{\footnotesize Non-diag}
& 2 &(80)56 &(80)56\\
& 3 & (1152)792 &(1152)792\\
& 4 & (11264)8024 &(11264)8024\\
& 5 & (89600)65984 &(89600)65984\\
& 6 & (626688)473768 &(626688)473768\\
\hline
\multirow{5}{*}{\footnotesize Diag}
& 2 &(40)40 & (10)10\\
& 3 &(320)306 & (98)76\\
& 4 & (1928)1776 & (533)383 \\
& 5 & (9008)8106 & (2779)1886\\
& 6 & (40128)35094 & (12893)8340\\
\hline
\end{tabular}
\end{table}

In the following, we compare the efficacy of the two digitization methods with respect to a specific Lorentzian field-theory observable: the energy correlators. These observables probe correlation functions of energy-flow operators which are examples of light-ray operators, and have been extensively studied in a wide range of collider experiments, establishing a compelling connection between collider phenomenology and formal quantum field theory~\cite{Moult:2025energyreview}. We consider the case by perturbing the vacuum $\lvert \Omega \rangle$ in the interaction theory with a source operator $O(0)$ at the origin and the corresponding sink operator $O(\bm{x})$ at position $\mathbf{x}$, the two-point energy–energy correlator is defined as
\begin{equation}\label{EEC}
\begin{aligned}
&\mathrm{EEC}(\theta)
\\
= &\int dt_1 \, dt_2 \,
\lim_{r_1,r_2 \to \infty}
r_1^{d-1} n_1^{i} \, r_2^{d-1} n_2^{j}
\\
&\times
\bigl\langle
\Omega \big|
O^{\dagger}(\bm x)\,
T_{0 i}(t_1, r_1 \mathbf{n}_1)\,
T_{0 j}(t_2, r_2 \mathbf{n}_2)\,
O(0)
\big|
\Omega
\bigr\rangle
\end{aligned}
\end{equation}
Here, $\mathbf{n}_1$ and $\mathbf{n}_2$ denote the two unit vectors specifying the directions of the observation points, and $\theta$ is the angle between them. The quantities $r_1$ and $r_2$ denote the distances of the first and second observation points from the origin. 

In the above, $T_{0i}(t, r \mathbf{n})$ is the energy-flux operator in the Heisenberg picture. It represents the energy flux at the space point $\mathbf{x} = r \mathbf{n}$ to its nearest-neighbor site $\mathbf{x}'$ in the $i$-th direction at time $t$. 
For notational clarity, we equivalently write $T_{0i}(t, r \mathbf{n})
\equiv
T_{0,\, \mathbf{x} \to \mathbf{x}'}(t)$. 

For the scalar field theory on the lattice, the energy-flux operator $T_{0,\, \mathbf{x} \to \mathbf{x}'}(t)$ can be constructed as follows. We decompose the Hamiltonian as a sum of local energy densities, with $H= \sum_{\bm{x}\in\Lambda} H(\bm x)$. In the Heisenberg picture, the time evolution of the local energy density is governed by 
\begin{equation}
    \partial_t H(\bm x, t) = -i [H, H(\bm x ,t)]=-i \sum_{\bm x'}[H(\bm x', t), H(\bm x, t)].
\end{equation}
where the sum runs over the nearest neighbors $\bm x'$ of $\bm x$. Yet, according to the energy conservation equation
\begin{equation}
    \partial_t H(\bm x, t) + \sum_{\bm x'} T_{0, \bm x\to \bm x'}(t) = 0
\end{equation}
Thus, we can identify the energy flux operator is
	\begin{equation}
    \begin{aligned}
		&T_{0,\bm x\to \bm x'}(t)\\&=-i [H(\bm x', t),H(\bm x, t)]\\&
        =e^{-i H t}\frac{1}{2}\delta x^{d-2}\bigg(\phi(\bm x)\Pi(\bm x')-\phi(\bm x')\Pi(\bm x)\bigg)e^{i H t}.
    \end{aligned}
    \label{eq:EnergyFlux}
	\end{equation}
In the following analysis, we consider a scalar field theory defined on a $2 \times 2$ spatial lattice. We will consider the OB with the occupation number truncated to $2^{n_q}$ and also the AB as adopted by JLP with its local Hilbert space truncated to $2^{n_q}$. We take the source and sink operators to be $O = \phi$, both inserted at the origin with coordinate ${\bm 0}=\{0, 0\}$. We first analyze the convergence behavior of
$T_{0,\{0,1\}\to\{1,1\}}(t)$ corresponding to the the energy flow from the lattice site $\{0,1\}$ to $\{1,1\}$ which is the observable $M(t)=\langle \Omega | \phi^\dagger({\bm 0})\, T_{0,\{0,1\}\to\{1,1\}}(t)\, \phi(\bm 0) | \Omega \rangle$, and then ${\rm EEC}(\theta)$, as functions of $n_q$. 
We then present an explicit implementation of the quantum circuits required for each component of the EEC calculation and provide numerical results for the energy flow 
$T_{0,\{0,1\}\to\{1,1\}}(t)$ using noiseless quantum simulations.

\subsection{Convergence of energy flow and EEC}
For each truncation $n_q$, we evaluate $M(t)$ using (i) the encoding scheme proposed in this work, and (ii) the AB-based scheme of JLP, as shown in \fig{l005}-\fig{l03}. 
Across these benchmarks, the results obtained with the OB encoding exhibit a faster convergence as $n_q$ is increased, while the curves from the AB basis at small $n_q$ can display a visibly larger deviation from the exact lattice result.

To quantify convergence at the level of an integrated energy-correlation observable, we calculate the corresponding two point energy-energy correlators on the lattice by considering different couplings. The two energy-flow directions are fixed along the lattice unit vectors, $\mathbf{n}_1 \equiv \mathbf{e}_1$ and $\mathbf{n}_2 \equiv \mathbf{e}_2$ which result in $\theta = \pi/2$. For the integration over time, we use time-discretized values computed over a fixed window $t \in [0,2]$ with step size $\delta t = 0.05$, so that
\begin{equation}
    \begin{aligned}
        	&{\rm{EEC_{lat}}}(\frac{\pi}{2})=(\delta t)^2 \sum_{t_1}\sum_{t_2}r_1^{d-1}r_2^{d-1}\\&\langle \Omega | \phi^\dagger({\bm 0})\, T_{0\mathbf{n}_1}(t_1, {\mathbf n_1} r_1)T_{0\mathbf{n}_2}(t_2, {\mathbf n_2} r_2) \phi({\bm 0}) | \Omega \rangle
    \end{aligned}
\end{equation}
Due to the periodic boundary conditions on the $2 \times 2$ lattice, the minimal distances $r_1$ and $r_2$ from the origin to lattice sites in the $\mathbf{e}_1$ and $\mathbf{e}_2$ directions are fixed at $\delta x$ for all times $t$ along the light cone. 

We observe that when the coupling is small, at $(m,\lambda)=(0.15,0.05)$, the OB results at $n_q= 2$ already converges to it $n_q=4$ value as shown in \tab{l005}, whereas the AB result at $n_q=2$ still exhibit a substantial discrepancy and requires a larger truncation to reach the same level of precision.
Similar convergence patterns persist for the larger couplings as shown in \tab{l015} and \tab{l03}, whereas the convergence of AB results are improved.
These results indicate that, for the energy-correlation observables considered here, the OB digitization can achieve a target precision with fewer qubits per local degree of freedom compared to the AB baseline of JLP.
\begin{figure}[t]
  \centering
  \input{fig_phidag_T_phi_intVEV}
  \caption{$\langle \Omega|\phi^\dagger(\bm 0)T_{0,\{0,1\}\to \{1,1\}}\phi(\bm 0)|\Omega\rangle$ as a function of the time $t$ for a $2\times2$ lattice with $m_0=0.15,\lambda=0.05$ and the renormalized mass $m = 0.20$. Comparison between exact lattice field calculation, the encoding scheme proposed in this work, and the encoding scheme of JLP.}
\label{fig:l005}
\end{figure}
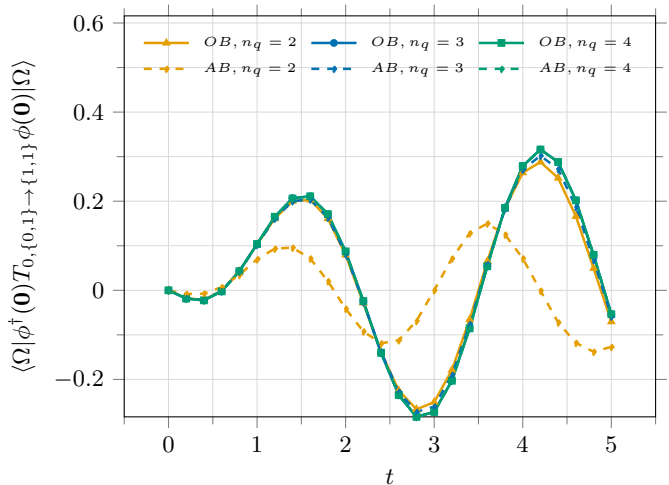
\begin{figure}[t]
  \centering
  \input{fig_phidag_T_phi_intVEV_m0.15lam0.15}
  \caption{Same plot as \fig{l005}, with $m_0=0.15,\lambda=0.15$ and the renormalized mass $m=0.26$.}
\label{fig:l015}
\end{figure}
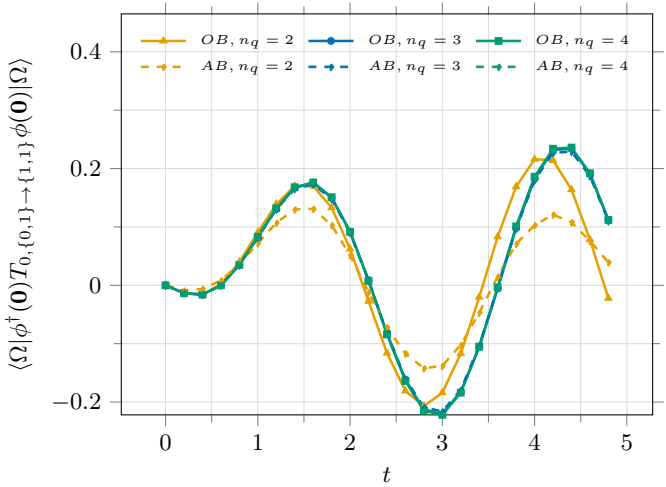
\begin{figure}[t]
  \centering
  \input{fig_phidag_T_phi_intVEV_m0.15lam0.3}
  \caption{Same plot as \fig{l005}, with $m_0=0.15,\lambda=0.3$ and the renormalized mass $m=0.32$.}
\label{fig:l03}
\end{figure}
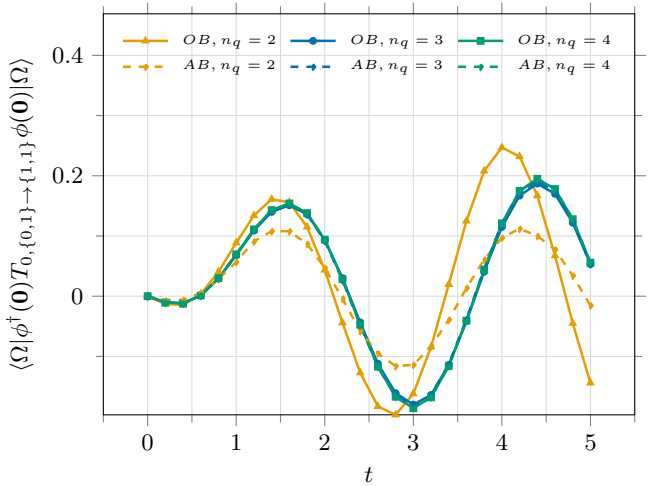
\begin{table}[htbp]
	\centering
	\caption{$\rm{EEC_{lat}}$ with $m_0=0.15, \lambda=0.05, t\in[0,2],dt=0.05$}
	\label{tab:l005}
	\begin{tabular}{lcr}
		\toprule
		$n_q$ & OB & AB \\
		\midrule
		2    & 0.2727    & 0.1281 \\
		3    & 0.2766   & 0.2698 \\
		4    & 0.2763   & 0.2762 \\
		\bottomrule
	\end{tabular}
\end{table}
\begin{table}[htbp]
	\centering
	\caption{$\rm{EEC_{lat}}$ with $m_0=0.15, \lambda=0.15, t\in[0,2],dt=0.05$ }
	\label{tab:l015}
	\begin{tabular}{lcr}
		\toprule
		$n_q$ & OB & AB \\
		\midrule
		2    & 0.1935    & 0.0999 \\
		3    & 0.1569   & 0.1538 \\
		4    & 0.1588   & 0.1572 \\
		\bottomrule
	\end{tabular}
\end{table}
\begin{table}[htbp]
	\centering
	\caption{$\rm{EEC_{lat}}$ with $m_0=0.15, \lambda=0.3, t\in[0,2],dt=0.05$ }
	\label{tab:l03}
	\begin{tabular}{lcr}
		\toprule
		$n_q$ & OB & AB \\
		\midrule
		2    & 0.1725    & 0.0812 \\
		3    & 0.0994   & 0.1026 \\
		4    & 0.1028   & 0.1028 \\
		\bottomrule
	\end{tabular}
\end{table}

\subsection{Quantum algorithms and resources}
In this section, we illustrate the quantum algorithms to calculate the energy flow and implement the quantum algorithms on a noiseless simulator to quantify the errors from trotterization. 
\subsubsection{Ground state preparation}
For a gapped system defined on a finite lattice, the interacting ground state $\ket{\Omega}$ can be prepared via adiabatic evolution by gradually turning on the interaction term. If the spectral gap between the ground state and first excited state remains finite along this interpolation path, the adiabatic theorem guarantees preparation of the interacting ground state with a runtime $T$ that scales inversely with minimum of the spectral gap $\Delta_{\min}$ along the path.

In lattice scalar field theory in $2+1$ dimensions, phase transitions occur at critical values of the bare parameters where the spectral gap vanishes in the continuum (infinite-volume) limit. At any finite lattice volume, the spectrum remains discrete and strictly gapped; however, near criticality the minimum gap can become parametrically small \cite{Jordan:2011ci, YeterAydeniz:2018sqft}, making adiabatic state preparation increasingly inefficient.

In the present work, we restrict our analysis to regions of parameter space that lie within the gapped phase and sufficiently far from the critical surface, so that the minimum gap $\Delta_{\min}$ remains finite. Ground-state preparation in or near critical regimes is deferred to variational quantum algorithms, whose optimized implementations for $2+1$d $\phi^4$ theory have been investigated in Ref.~\cite{Cao:2025shc}.

The adiabatic state preparation is implemented by simulating the time-dependent Hamiltonian
\begin{equation}
    H_{\rm ad}(t) = H_0 + \frac{t}{T} H_{\rm int},
\end{equation}
which linearly interpolates between $H_0$ and $H_0+H_{\rm int}$. The initial state is taken to be the ground state of $H_0$, namely the free vacuum annihilated by all $a_{\bm p}$. The time evolution operator corresponding to $H_{\rm int}$ is thus implemented with the diagonalization method.

\subsubsection{Quantum circuits for measuring Lorentzian observables}
After preparing the interacting ground state $|\Omega\rangle$, we compute physical observables by evaluating expectation values in this state. For example, one may extract time-dependent quantities such as $M(t)$, or correlation functions of the form
\[
\langle \Omega | \phi^{\dagger}(\mathbf{0})\,
T_{0 i}(t_1, \mathbf{n}_1 r_1)\,
T_{0 j}(t_2, \mathbf{n}_2 r_2)\,
\phi(\mathbf{0}) | \Omega \rangle,
\]
which enter, for instance, in the calculation of the energy--energy correlator ${\rm EEC}_{\rm latt}(\theta)$.

More generally, the observables of interest are expectation values of operators $W$ in the interacting ground state,
\[
\mathrm{Obs} = \langle \Omega | W | \Omega \rangle.
\]
If $W$ is unitary, its real and imaginary parts can be obtained via the standard Hadamard test \cite{Lin:2022vrd}. An ancilla qubit is initialized in $|0\rangle$, followed by a Hadamard gate, a controlled-$W$ operation, and a second Hadamard gate prior to measurement. The probability of measuring the ancilla in the $|0\rangle$ state is $(1+\operatorname{Re}(\text{Obs}))/2$. To extract the imaginary part, an additional $S^\dagger$ gate is applied to the ancilla before the final Hadamard gate, yielding $(1+\operatorname{Im}(\text{Obs}))/2$. In our case, operators such as $\phi(\bm 0)$ and 
$T_{0,\bm x\to \bm x'}(t)$ are Hermitian but generally non-unitary, which
must be embedded into a unitary operator in order to apply the Hadamard test.

Given \eq{EnergyFlux}, the time dependent part can be implemented with trotterization in the diagonalization framework, then it is sufficient to implement only the operators $\phi(\bm x)$ and $\Pi(\bm x)$ on the quantum circuit as unitary operators which can be achieved using Linear Combination of Unitaries (LCU) method. 
The number of ancilla qubits required in the LCU method  
depends on the total number of Pauli strings appearing in the operator decomposition.  
Since we have diagonalized each subspace of size $n_{q}$ qubits, the number of independent  
Pauli strings in a single subspace is $2^{n_{q}}$. Consequently, the total number of Pauli  
strings in the full system of size $N^{d}$ is at most $N^{d} \cdot 2^{n_{q}}$. Therefore, the number of required ancilla qubits in the LCU method to implement a single $\phi(\bm x)$ or $\Pi(\bm x)$ is given by
\begin{equation}
    N_{\mathrm{LCU}} =\lceil \log_{2}\!\bigl(N^{d} \cdot 2^{n_{q}}\bigr)\rceil=\lceil d \log_{2} N + n_{q}\rceil.
\end{equation}

As for the energy flux operator involving multiplications of Hermitian operator $\phi(\bm x)$ and $\Pi(\bm x)$, the Multiple Coherent Measurement (MCM) circuit as proposed in \cite{Vasconcelos:2025mwu} can be used to reduce the number of ancilla qubits in LCU. It is proved that $N_{\rm MCM} = \lceil\log_2 K\rceil$ ancillae is optimal, where $K$ is the number of the operators in the multiplication.

The core component of the MCM circuit is the unitary block $U_i$ for the $i$-th operator in the multiplication, with $U_i$ consists of two subcircuits: 
\begin{enumerate}
    \item The first subcircuit acts on $N_{\mathrm{MCM}}$ ancilla qubits, 
    mapping the initial state $\lvert 0^{N_{\mathrm{MCM}}}\rangle$ 
    to the computational basis state $\lvert i \rangle$ 
    (with $i = 0,1,\ldots,K $). 
    
    \item The second subcircuit acts jointly on the 
    $N_{\mathrm{LCU}}+N_{\mathrm{MCM}}$ ancilla qubits, 
    and implements a controlled operation: 
    whenever the $N_{\mathrm{LCU}}$ ancillas are in the state 
    $\lvert 0^{N_{\mathrm{LCU}}}\rangle$, 
    the state of the $N_{\mathrm{MCM}}$ qubits is mapped back from 
    $\lvert i \rangle$ to $\lvert 0^{N_{\mathrm{MCM}}}\rangle$. 
\end{enumerate}

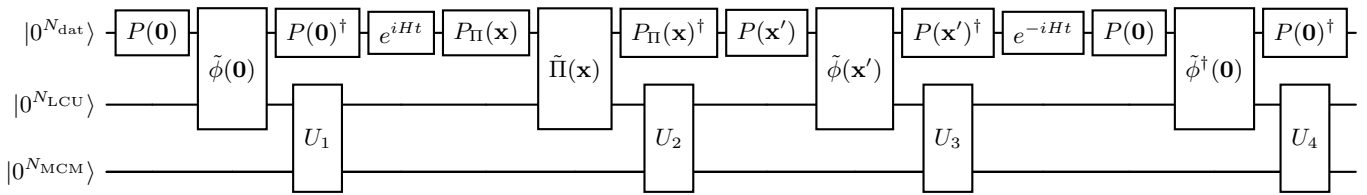
\begin{figure*}[ht]
\centering
\begin{quantikz}[row sep=0.3cm, column sep=0.12cm]
\lstick{$|0^{N_{\mathrm{dat}}}\rangle$} &\gate{P(\bm 0)} & \gate[wires=2]{\tilde{\phi}(\bm 0)} & \gate{P(\bm 0)^\dagger} & \gate{e^{i H t}} &\gate{P_\Pi(\bm x)} & \gate[wires=2]{\tilde{\Pi}(\bm x)} & \gate{P_\Pi(\bm x)^\dagger}&           \gate{P(\bm x')} &\gate[wires=2]{\tilde{\phi}(\bm x')} &\gate{P(\bm x')^\dagger} &\gate{e^{-i H t}} & \gate{P(\bm 0)} &\gate[wires=2]{\tilde{\phi}^\dagger(\bm 0)} &\gate{P(\bm 0)^\dagger} & \\
\lstick{|$0^{N_{\mathrm{LCU}}}\rangle$} &  & &  \gate[wires=2]{U_1}& && &\gate[wires=2]{U_2} &   & &\gate[wires=2]{U_3} &&&&\gate[wires=2]{U_4}&\\
\lstick{$|0^{N_{\mathrm{MCM}}}\rangle$} & \qw & \qw & \qw & \qw &\qw &\qw& \qw &\qw &\qw&\qw&\qw&\qw&\qw&\qw &
\end{quantikz}
\caption{Realization of operator $\phi^\dagger(\bm 0) e^{-i H t}\phi(\bm x')\Pi(\bm x)e^{i H t}\phi(\bm 0)$ as quantum circuit.}
\label{fig:circuit}
\end{figure*}

This construction ensures that, when all $N_{\rm LCU} + N_{\rm MCM}$ ancilla registers are in the zero state, 
the effective action on the data register corresponds exactly to the desired 
product of operators. A demonstration for the circuit illustrating the quantum circuit implementing $\phi^\dagger(\bm 0) e^{-i H t}\phi(\bm x')\Pi(\bm x)e^{i H t}\phi(\bm 0)$ is shown in \fig{circuit}.
Using the LCU and MCM methods, the total number of ancilla qubits to calculate ${\rm EEC}(\theta)$ is
\begin{equation}
\begin{aligned}
    N_{\mathrm{anc}} &= N_{\mathrm{LCU}} + N_{\mathrm{MCM}} \\
    &=\lceil d \log_2 N + n_{q} \rceil+ \lceil\log_2 K\rceil.
\end{aligned}
\end{equation}
with $K = 6$.

We now turn to the issue of circuit depth.  
If there are $K$ numbers of $\phi(\bm x)$ or $\Pi(\bm x)$ in the multiplication, a total of $2K$ Hermite transformations are required. 
The total circuit depth for implementing the $2K$ transformation matrices is estimated as
\begin{equation}
D_{\mathrm{P}} \sim \mathcal{O}\bigl(2K \cdot 4^{n_q}\bigr) \sim \mathcal{O}\bigl(K\cdot 2^{2n_q+1}\bigr).
\end{equation}

Since the diagonalized operators $\phi(\bm{x})$ and $\Pi(\bm{x})$ decompose into a total of $N^d \times 2^{n_q}$ Pauli strings, with no known structure in their coefficients, the circuit depth of the LCU implementation scales as
\[
D_{\rm LCU,\,single} \sim \mathcal{O}(N^d \, 2^{n_q}) \, ,
\]
where each individual Pauli string can be implemented with circuit depth $\mathcal{O}(1)$.
If we wish to compute the product of $K$ operators, the total depth required for the LCU circuits is  
\begin{equation}
D_{\mathrm{LCU}} \sim 
\mathcal{O}\bigl(K N^d 2^{n_q}\bigr).
\end{equation}

The implementation of the MCM method relies on the matrices $U_i$.  
Each matrix $U_i$ acts on the ancilla register by mapping the all-zero state 
$|0^{\otimes}\rangle$ to the basis state $|i\rangle$, and subsequently 
mapping the desired state back to $|0^{\otimes}\rangle$.  
The circuit depth required for this process is estimated as
\begin{equation}
\begin{aligned}
D_{\mathrm{MCM,single}} &\sim \mathcal{O}(N_{\mathrm{LCU}} + N_{\mathrm{MCM}}).
\end{aligned}
\end{equation}
For the product of $K$ operators, $K$ unitary matrices $U_i$ are required,  
and the corresponding circuit depth is
\begin{equation}
D_{\mathrm{MCM}}  \sim \mathcal{O}(K N_{\rm LCU} + K N_{\rm MCM}).
\end{equation}

\subsubsection{Numerical results}
\begin{figure}[h]
  \centering
  \input{fig_numerical_phiTphi_inter}  
  \caption{$\langle \Omega|\phi^\dagger(\bm 0)T_{0,\{0,1\}\to \{1,1\}}\phi(\bm 0)|\Omega\rangle$ as a function of the time $t$ for a $2\times2$ lattice
with $\delta x$ = 1, $n_q=2$ and $m_0 = 0.15$ and $\lambda = 0.05$. }
  \label{fig:numerical_phiTphi}
\end{figure}
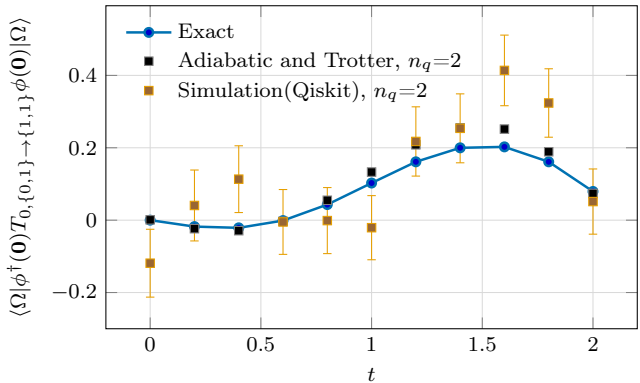
We construct the quantum circuit following the algorithm described above and perform numerical simulations. As a representative example, we evaluate the expectation value
\[
\langle \Omega \vert \phi^{\dagger}(\bm{0}) \, T_{0,\{0,1\}\to \{1,1\}}(t) \, \phi(\bm{0}) \vert \Omega \rangle \, .
\]
The initial state is prepared via adiabatic evolution with total evolution time $T = 3$ and time step $\delta t = 0.05$, which results in a total circuit depth of $\mathcal{O}(10^5)$. Using this set of parameters, the prepared ground state $\ket{\Omega}$ achieves a fidelity of $0.99$. 

Real-time evolution is implemented using a Trotter decomposition with time step $\delta t = 0.05$, where each Trotter step has a circuit depth comparable to that of the adiabatic evolution. The additional circuit depth required to implement the $\phi$ and $\Pi$ operators in the energy-flux measurement is $\mathcal{O}(10^6)$.

The results are shown in \fig{numerical_phiTphi}. The blue dotted line corresponds to the exact result. The black points are obtained from exact matrix simulations of the quantum circuit, incorporating the Trotterization errors in both adiabatic evolution and real-time evolution. As shown in \fig{numerical_phiTphi}, the Trotter error increases with evolution time, but remains below $20\%$ for $t \leq 2$. Note that the Trotter
error here arises from the non-commutativity between $H_0$
and $H_{\text{int}}$ when approximating the time-evolution operator
associated with the full Hamiltonian $H$. In contrast, the
time evolution generated by $H_{\text{int}}$ is implemented exactly
using the diagonalization-based method.The yellow points correspond to results obtained from the Qiskit simulator, where each data point is averaged over 28 independent runs, each consisting of $10^5$ measurement shots. The error bars quantify the statistical uncertainty (one standard deviation) computed from the distribution over these 28 repetitions, and the results are consistent with classical simulations within $2\sigma$.
\section{Conclusions}
\label{sec:conclusion}
In this work, we have developed improved quantum algorithms for simulating interacting scalar quantum field theories using occupation-basis (OB) digitization. By diagonalizing the truncated field operator prior to Pauli decomposition, we showed that the interaction evolution operator can be implemented using only commuting $Z$-type Pauli strings. This eliminates the large Trotter errors present in direct Pauli-string constructions and significantly reduces circuit depth.

We quantified these improvements by explicitly constructing quantum circuits and comparing CNOT depths for quartic interactions. For representative lattice sizes and truncations, the diagonalization-based method yields substantial reductions in circuit depth—by up to several orders of magnitude, in comparison to the existing approaches based on direct decomposition into Pauli strings.

As a physics benchmark, we evaluated Lorentzian energy--energy correlators on a $2\times 2$ lattice. Our results show that OB digitization can exhibit faster convergence with respect to the local truncation size compared to the amplitude-basis (AB) approach of Jordan, Lee, and Preskill. In particular, for weak to moderate couplings, accurate results can be achieved with fewer qubits per site, highlighting the efficiency of OB encoding for real-time observables in the NISQ era.

Finally, using noiseless quantum simulations, we assessed the combined impact of adiabatic state preparation, Trotterization, and measurement statistics. We find that the overall errors remain controlled within the time window studied, with trotterization errors below $20\%$, and statistical uncertainties well controlled with a large number of measurement shots. 

Taken together, these results establish diagonalization-based OB digitization as a promising route toward efficient quantum simulations of real-time quantum field theory in the NISQ era, offering both algorithmic advances and phenomenological benchmarks for future studies of light-ray observables on near-term quantum devices.
\section*{Acknowledgement}
We would like to thank Xingyu Guo, Danbo Zhang for useful discussions.
This work is partially supported by the National Natural Science Foundation of China under Grant Nos. 12522509 and 12235001, and by the National Key R\&D Program of China under Contract No. 2025YFA1614200. X.L. is supported by by the National Natural Science Foundation of China under Grant Nos. 12547109 and Fundamental Research Funds for the Central Universities, Beijing
Normal University. This work was also supported by the Fundamental and Interdisciplinary Disciplines Breakthrough Plan of the Ministry of Education of China-JYB2025XDXM204. The authors gratefully acknowledge the valuable discussions and insights provided by the members of the Collaboration on Precision Tests and New Physics (CPTNP).
\bibliography{ref}
\clearpage
\appendix
\end{document}

%% file: preamble.tex
\usepackage{mathtools}
\usepackage{simplewick}
\usepackage{amsmath,amssymb,amsbsy,amsfonts}
\usepackage{bm}
\let\oldbm\bm

\renewcommand{\bm}[1]{%
  \ifcat\noexpand#1\relax
    \oldbm{#1}%
  \else
    \mathbf{#1}%
  \fi
}
\usepackage{graphics}
\graphicspath{{./fig/}}

\usepackage{ninecolors}
\usepackage[normalem]{ulem}

\usepackage{tikz}
\usepackage{tikz-feynman} 
\tikzfeynmanset{compat=1.1.0}
\usetikzlibrary{3d}
\usepackage{pgfplots}
\pgfplotsset{compat=1.18}
\usepgfplotslibrary{statistics}
\usepackage{pgfplotstable}

\newcommand{\eq}[1]{Eq.~(\ref{eq:#1})}
\newcommand{\tab}[1]{Tab.~(\ref{tab:#1})}
\newcommand{\fig}[1]{Fig.~(\ref{fig:#1})}

\usepackage{microtype}

\usepackage{physics}
\usepackage{physics2}
\usephysicsmodule{ab,ab.braket}

\newcommand\del\partial

\usepackage{dutchcal}

\usepackage{xcolor}

\usepackage{makecell}

\usepackage{enumitem}
\usepackage{pifont}

\usepackage{graphicx}
\usepackage{amsfonts}
\usepackage{tocvsec2}
\usepackage{slashed}
\usepackage{cancel}
\usepackage{comment}

\usepackage{minibox}
\usepackage{orcidlink}
\usepackage{lipsum}

\newcommand\beq{\begin{eqnarray}}
\newcommand\eeq{\end{eqnarray}}

\newcommand{\mybar}[1]{\kern 0.6pt\overline{\kern -0.6pt#1\kern -0.6pt}\kern 0.6pt}
\usepackage{ulem}

\def\U\Omega{U(1)_{\Omega}}

\usepackage[section]{placeins}

\usepackage{xparse}

\NewDocumentCommand{\Dw}{o}{%
  \IfValueTF{#1}{%
    D^{(#1)}_w
  }{%
    D_w
  }
}

\usepackage{bbold}
\usepackage{slashed}
\usepackage{braket}

\usepackage[capitalise]{cleveref}
\crefname{section}{Sec.}{Secs.}

\usepackage{hyperref}
\hypersetup{
    unicode,
    pdfprintscaling=None,
    colorlinks,
    breaklinks
}
\hypersetup{
  colorlinks=true,   
  linkcolor=cyan,    
  citecolor=blue,    
  filecolor=magenta, 
  urlcolor=blue      
}


%% file: fig_phidag_T_phi_intVEV.tex
\begin{tikzpicture}
  \definecolor{TolBlue}{RGB}{0,114,178}   
  \definecolor{TolOrange}{RGB}{230,159,0} 
  \definecolor{TolGreen}{RGB}{0,158,115}  

  \begin{axis}[
    width=\columnwidth, height=0.8\columnwidth,
    xlabel={$t$}, ylabel={$\langle\Omega|\phi^\dagger(\bm 0)T_{0,\{0,1\}\to \{1,1\}}\phi(\bm 0)|\Omega\rangle$},
    tick align=outside, tick style={thin}, minor tick num=1,
    scaled ticks=false,
    enlarge y limits={upper,value=0.5},
    legend style={
  at={(0.5,1.0)}, anchor=north, yshift=-1.2ex,
  draw=none, fill=none,
  font=\tiny, legend columns=3, column sep=3pt, inner sep=1pt,row sep=1pt
},
    legend cell align=left,
    unbounded coords=discard, clip=true,
    grid=both, grid style={line width=0.2pt, draw=black!15},
    clip=false
  ]

    \pgfplotstableread[row sep=newline, col sep=space]{
t	OBnq3	OBnq4	OBnq2	OBnq1	ABnq4	ABnq3	ABnq2	ABnq1
0	4.86E-16	-6.96E-16	-4.93E-16	4.04E-14	-5.72E-10	5.01E-06	-3.19E-04	-7.06E-04
0.2	-1.86E-02	-1.85E-02	-1.78E-02	4.04E-14	-1.85E-02	-1.77E-02	-8.19E-03	-3.56E-03
0.4	-2.25E-02	-2.25E-02	-2.12E-02	4.05E-14	-2.24E-02	-2.13E-02	-7.41E-03	-8.04E-03
0.6	-2.23E-03	-2.23E-03	-1.14E-03	4.05E-14	-2.23E-03	-1.67E-03	7.04E-03	8.39E-04
0.8	4.30E-02	4.29E-02	4.31E-02	4.05E-14	4.29E-02	4.22E-02	3.47E-02	3.28E-02
1	1.04E-01	1.04E-01	1.03E-01	4.05E-14	1.04E-01	1.02E-01	6.97E-02	5.99E-02
1.2	1.65E-01	1.65E-01	1.61E-01	4.05E-14	1.65E-01	1.61E-01	9.36E-02	3.75E-02
1.4	2.07E-01	2.06E-01	2.00E-01	4.06E-14	2.06E-01	2.00E-01	9.58E-02	-2.18E-02
1.6	2.11E-01	2.11E-01	2.03E-01	4.06E-14	2.11E-01	2.04E-01	7.15E-02	-4.20E-02
1.8	1.71E-01	1.71E-01	1.61E-01	4.07E-14	1.71E-01	1.63E-01	2.05E-02	1.72E-02
2	8.72E-02	8.73E-02	7.93E-02	4.07E-14	8.73E-02	8.17E-02	-4.14E-02	9.63E-02
2.2	-2.43E-02	-2.41E-02	-2.83E-02	4.07E-14	-2.41E-02	-2.65E-02	-9.22E-02	1.10E-01
2.4	-1.41E-01	-1.40E-01	-1.39E-01	4.07E-14	-1.40E-01	-1.38E-01	-1.19E-01	5.10E-02
2.6	-2.35E-01	-2.35E-01	-2.25E-01	4.08E-14	-2.35E-01	-2.28E-01	-1.12E-01	-1.74E-02
2.8	-2.84E-01	-2.84E-01	-2.67E-01	4.08E-14	-2.84E-01	-2.73E-01	-6.87E-02	-3.79E-02
3	-2.73E-01	-2.73E-01	-2.51E-01	4.08E-14	-2.73E-01	-2.61E-01	2.05E-04	-3.27E-03
3.2	-2.02E-01	-2.03E-01	-1.79E-01	4.08E-14	-2.03E-01	-1.90E-01	7.14E-02	5.17E-02
3.4	-8.49E-02	-8.54E-02	-6.56E-02	4.09E-14	-8.53E-02	-7.58E-02	1.28E-01	7.33E-02
3.6	5.45E-02	5.40E-02	6.52E-02	4.09E-14	5.40E-02	5.85E-02	1.50E-01	3.19E-02
3.8	1.86E-01	1.85E-01	1.84E-01	4.09E-14	1.85E-01	1.83E-01	1.25E-01	-3.68E-02
4	2.79E-01	2.79E-01	2.64E-01	4.09E-14	2.79E-01	2.71E-01	7.16E-02	-6.32E-02
4.2	3.16E-01	3.16E-01	2.88E-01	4.10E-14	3.16E-01	3.02E-01	-8.76E-04	-3.07E-02
4.4	2.88E-01	2.88E-01	2.52E-01	4.10E-14	2.88E-01	2.72E-01	-7.22E-02	6.60E-03
4.6	2.02E-01	2.02E-01	1.66E-01	4.11E-14	2.02E-01	1.87E-01	-1.18E-01	6.67E-03
4.8	7.89E-02	7.95E-02	4.94E-02	4.11E-14	7.95E-02	6.82E-02	-1.38E-01	-7.57E-03
5	-5.42E-02	-5.35E-02	-7.04E-02	4.11E-14	-5.35E-02	-5.89E-02	-1.27E-01	-7.11E-04
    }\datatable

    \addplot[mark=triangle*,mark size=1pt,line width=0.95pt, color=TolOrange] table[x=t, y=OBnq2] {\datatable};
    \addlegendentry{$OB,n_q=2$}
    \addplot[mark=*, mark size=1pt, line width=0.95pt, color=TolBlue] table[x=t, y=OBnq3] {\datatable};
    \addlegendentry{$OB,n_q=3$}
    \addplot[mark=square*,mark size=1pt,line width=0.95pt, color=TolGreen] table[x=t, y=OBnq4] {\datatable};
    \addlegendentry{$OB,n_q=4$}

    \addplot[mark=diamond*, dashed,mark size=1pt,line width=0.95pt, color=TolOrange] table[x=t, y=ABnq2] {\datatable};
    \addlegendentry{$AB,n_q=2$}
    \addplot[mark=diamond*, dashed,mark size=1pt,line width=0.95pt, color=TolBlue] table[x=t, y=ABnq3] {\datatable};
    \addlegendentry{$AB,n_q=3$}
    \addplot[mark=diamond*, dashed,mark size=1pt,line width=0.95pt, color=TolGreen] table[x=t, y=ABnq4] {\datatable};
    \addlegendentry{$AB,n_q=4$}
\end{axis}
\end{tikzpicture}

%% file: fig_phidag_T_phi_intVEV_m0.15lam0.15.tex
\begin{tikzpicture}
  \definecolor{TolBlue}{RGB}{0,114,178}   
  \definecolor{TolOrange}{RGB}{230,159,0} 
  \definecolor{TolGreen}{RGB}{0,158,115}  

  \begin{axis}[
    width=\columnwidth, height=0.8\columnwidth,
    xlabel={$t$}, ylabel={$\langle\Omega|\phi^\dagger(\bm 0)T_{0,\{0,1\}\to \{1,1\}}\phi(\bm 0)|\Omega\rangle$},
    tick align=outside, tick style={thin}, minor tick num=1,
    scaled ticks=false,
    enlarge y limits={upper,value=0.5},
    legend style={
  at={(0.5,1.0)}, anchor=north, yshift=-1.2ex,
  draw=none, fill=none,
  font=\tiny, legend columns=3, column sep=3pt, inner sep=1pt,row sep=1pt
},
    legend cell align=left,
    unbounded coords=discard, clip=true,
    grid=both, grid style={line width=0.2pt, draw=black!15},
    clip=false
  ]

    \pgfplotstableread[row sep=newline, col sep=space]{
t	OBnq3	OBnq4	OBnq2	OBnq1	ABnq4	ABnq3	ABnq2	ABnq1
0	-8.92E-16	2.11E-15	-5.66E-16	4.04E-14	-1.27E-07	3.62E-06	-1.79E-04	-7.06E-04
0.2	-1.34E-02	-1.35E-02	-1.42E-02	4.04E-14	-1.34E-02	-1.31E-02	-9.15E-03	-3.56E-03
0.4	-1.58E-02	-1.60E-02	-1.61E-02	4.05E-14	-1.58E-02	-1.54E-02	-6.16E-03	-8.04E-03
0.6	-7.90E-05	-1.36E-04	1.84E-03	4.05E-14	-1.22E-05	1.52E-04	8.38E-03	8.39E-04
0.8	3.47E-02	3.50E-02	4.01E-02	4.05E-14	3.49E-02	3.46E-02	3.60E-02	3.28E-02
1	8.25E-02	8.31E-02	9.06E-02	4.05E-14	8.28E-02	8.18E-02	7.22E-02	5.99E-02
1.2	1.31E-01	1.32E-01	1.39E-01	4.05E-14	1.32E-01	1.30E-01	1.07E-01	3.75E-02
1.4	1.67E-01	1.68E-01	1.70E-01	4.06E-14	1.67E-01	1.65E-01	1.30E-01	-2.18E-02
1.6	1.76E-01	1.76E-01	1.70E-01	4.06E-14	1.76E-01	1.73E-01	1.31E-01	-4.20E-02
1.8	1.50E-01	1.51E-01	1.33E-01	4.07E-14	1.50E-01	1.48E-01	1.03E-01	1.72E-02
2	9.18E-02	9.16E-02	6.23E-02	4.07E-14	9.13E-02	8.98E-02	5.16E-02	9.63E-02
2.2	8.80E-03	7.86E-03	-2.72E-02	4.07E-14	8.10E-03	7.94E-03	-1.14E-02	1.10E-01
2.4	-8.24E-02	-8.40E-02	-1.16E-01	4.07E-14	-8.31E-02	-8.17E-02	-7.26E-02	5.10E-02
2.6	-1.62E-01	-1.64E-01	-1.81E-01	4.08E-14	-1.63E-01	-1.60E-01	-1.17E-01	-1.74E-02
2.8	-2.13E-01	-2.15E-01	-2.07E-01	4.08E-14	-2.13E-01	-2.09E-01	-1.42E-01	-3.79E-02
3	-2.21E-01	-2.22E-01	-1.84E-01	4.08E-14	-2.20E-01	-2.16E-01	-1.38E-01	-3.27E-03
3.2	-1.83E-01	-1.84E-01	-1.17E-01	4.08E-14	-1.82E-01	-1.79E-01	-1.02E-01	5.17E-02
3.4	-1.06E-01	-1.05E-01	-2.01E-02	4.09E-14	-1.05E-01	-1.03E-01	-4.69E-02	7.33E-02
3.6	-5.31E-03	-3.63E-03	8.32E-02	4.09E-14	-4.31E-03	-4.31E-03	1.49E-02	3.19E-02
3.8	9.84E-02	1.01E-01	1.69E-01	4.09E-14	9.93E-02	9.70E-02	7.10E-02	-3.68E-02
4	1.83E-01	1.86E-01	2.16E-01	4.09E-14	1.84E-01	1.80E-01	1.03E-01	-6.32E-02
4.2	2.32E-01	2.34E-01	2.14E-01	4.10E-14	2.32E-01	2.27E-01	1.21E-01	-3.07E-02
4.4	2.34E-01	2.36E-01	1.64E-01	4.10E-14	2.34E-01	2.29E-01	1.08E-01	6.60E-03
4.6	1.91E-01	1.92E-01	7.83E-02	4.11E-14	1.91E-01	1.87E-01	7.61E-02	6.67E-03
4.8	1.12E-01	1.12E-01	-2.22E-02	4.11E-14	1.12E-01	1.10E-01	3.94E-02	-7.57E-03
    }\datatable
    \addplot[mark=triangle*,mark size=1pt,line width=0.95pt, color=TolOrange] table[x=t, y=OBnq2] {\datatable};
    \addlegendentry{$OB,n_q=2$}
    \addplot[mark=*, mark size=1pt, line width=0.95pt, color=TolBlue] table[x=t, y=OBnq3] {\datatable};
    \addlegendentry{$OB,n_q=3$}
    \addplot[mark=square*,mark size=1pt,line width=0.95pt, color=TolGreen] table[x=t, y=OBnq4] {\datatable};
    \addlegendentry{$OB,n_q=4$}

    \addplot[mark=diamond*, dashed,mark size=1pt,line width=0.95pt, color=TolOrange] table[x=t, y=ABnq2] {\datatable};
    \addlegendentry{$AB,n_q=2$}
    \addplot[mark=diamond*, dashed,mark size=1pt,line width=0.95pt, color=TolBlue] table[x=t, y=ABnq3] {\datatable};
    \addlegendentry{$AB,n_q=3$}
    \addplot[mark=diamond*, dashed,mark size=1pt,line width=0.95pt, color=TolGreen] table[x=t, y=ABnq4] {\datatable};
    \addlegendentry{$AB,n_q=4$}
\end{axis}
\end{tikzpicture}

%% file: fig_phidag_T_phi_intVEV_m0.15lam0.3.tex
\begin{tikzpicture}
  \definecolor{TolBlue}{RGB}{0,114,178}   
  \definecolor{TolOrange}{RGB}{230,159,0} 
  \definecolor{TolGreen}{RGB}{0,158,115}  

  \begin{axis}[
    width=\columnwidth, height=0.8\columnwidth,
    xlabel={$t$}, ylabel={$\langle\Omega|\phi^\dagger(\bm 0)T_{0,\{0,1\}\to \{1,1\}}\phi(\bm 0)|\Omega\rangle$},
    tick align=outside, tick style={thin}, minor tick num=1,
    scaled ticks=false,
    enlarge y limits={upper,value=0.5},
    legend style={
  at={(0.5,1.0)}, anchor=north, yshift=-1.2ex,
  draw=none, fill=none,
  font=\tiny, legend columns=3, column sep=3pt, inner sep=1pt,row sep=1pt
},
    legend cell align=left,
    unbounded coords=discard, clip=true,
    grid=both, grid style={line width=0.2pt, draw=black!15},
    clip=false
  ]

    \pgfplotstableread[row sep=newline, col sep=space]{
t	OBnq3	OBnq4	OBnq2	OBnq1	ABnq4	ABnq3	ABnq2	ABnq1
0	9.54E-17	-8.05E-16	2.81E-16	4.04E-14	-2.06E-09	1.48E-06	-1.55E-05	-7.06E-04
0.2	-9.99E-03	-1.04E-02	-1.31E-02	4.04E-14	-1.04E-02	-1.02E-02	-6.97E-03	-3.56E-03
0.4	-1.15E-02	-1.20E-02	-1.45E-02	4.05E-14	-1.20E-02	-1.18E-02	-6.07E-03	-8.04E-03
0.6	1.32E-03	1.12E-03	3.42E-03	4.05E-14	1.12E-03	1.19E-03	5.66E-03	8.39E-04
0.8	2.93E-02	2.99E-02	4.05E-02	4.05E-14	2.99E-02	2.97E-02	2.89E-02	3.28E-02
1	6.80E-02	6.96E-02	8.87E-02	4.05E-14	6.96E-02	6.93E-02	5.69E-02	5.99E-02
1.2	1.09E-01	1.11E-01	1.34E-01	4.05E-14	1.11E-01	1.10E-01	9.11E-02	3.75E-02
1.4	1.40E-01	1.43E-01	1.61E-01	4.06E-14	1.43E-01	1.42E-01	1.08E-01	-2.18E-02
1.6	1.51E-01	1.54E-01	1.56E-01	4.06E-14	1.54E-01	1.53E-01	1.08E-01	-4.20E-02
1.8	1.36E-01	1.38E-01	1.15E-01	4.07E-14	1.38E-01	1.36E-01	8.69E-02	1.72E-02
2	9.35E-02	9.39E-02	4.36E-02	4.07E-14	9.40E-02	9.19E-02	4.70E-02	9.63E-02
2.2	3.00E-02	2.83E-02	-4.40E-02	4.07E-14	2.84E-02	2.65E-02	-3.85E-03	1.10E-01
2.4	-4.34E-02	-4.71E-02	-1.27E-01	4.07E-14	-4.69E-02	-4.81E-02	-5.79E-02	5.10E-02
2.6	-1.12E-01	-1.17E-01	-1.83E-01	4.08E-14	-1.17E-01	-1.17E-01	-9.53E-02	-1.74E-02
2.8	-1.61E-01	-1.67E-01	-1.97E-01	4.08E-14	-1.67E-01	-1.66E-01	-1.16E-01	-3.79E-02
3	-1.80E-01	-1.86E-01	-1.62E-01	4.08E-14	-1.86E-01	-1.83E-01	-1.14E-01	-3.27E-03
3.2	-1.64E-01	-1.68E-01	-8.43E-02	4.08E-14	-1.68E-01	-1.64E-01	-8.36E-02	5.17E-02
3.4	-1.14E-01	-1.16E-01	1.96E-02	4.09E-14	-1.16E-01	-1.13E-01	-3.91E-02	7.33E-02
3.6	-4.12E-02	-4.05E-02	1.25E-01	4.09E-14	-4.07E-02	-3.79E-02	1.37E-02	3.19E-02
3.8	4.07E-02	4.42E-02	2.08E-01	4.09E-14	4.39E-02	4.54E-02	6.08E-02	-3.68E-02
4	1.15E-01	1.21E-01	2.47E-01	4.09E-14	1.21E-01	1.21E-01	9.71E-02	-6.32E-02
4.2	1.67E-01	1.75E-01	2.32E-01	4.10E-14	1.74E-01	1.73E-01	1.12E-01	-3.07E-02
4.4	1.87E-01	1.95E-01	1.67E-01	4.10E-14	1.95E-01	1.91E-01	1.00E-01	6.60E-03
4.6	1.70E-01	1.78E-01	6.73E-02	4.11E-14	1.78E-01	1.74E-01	7.76E-02	6.67E-03
4.8	1.22E-01	1.28E-01	-4.49E-02	4.11E-14	1.28E-01	1.24E-01	3.41E-02	-7.57E-03
5	5.32E-02	5.57E-02	-1.44E-01	4.11E-14	5.60E-02	5.31E-02	-1.55E-02	-7.11E-04
    }\datatable

    \addplot[mark=triangle*,mark size=1pt,line width=0.95pt, color=TolOrange] table[x=t, y=OBnq2] {\datatable};
    \addlegendentry{$OB,n_q=2$}
    \addplot[mark=*, mark size=1pt, line width=0.95pt, color=TolBlue] table[x=t, y=OBnq3] {\datatable};
    \addlegendentry{$OB,n_q=3$}
    \addplot[mark=square*,mark size=1pt,line width=0.95pt, color=TolGreen] table[x=t, y=OBnq4] {\datatable};
    \addlegendentry{$OB,n_q=4$}

    \addplot[mark=diamond*, dashed,mark size=1pt,line width=0.95pt, color=TolOrange] table[x=t, y=ABnq2] {\datatable};
    \addlegendentry{$AB,n_q=2$}
    \addplot[mark=diamond*, dashed,mark size=1pt,line width=0.95pt, color=TolBlue] table[x=t, y=ABnq3] {\datatable};
    \addlegendentry{$AB,n_q=3$}
    \addplot[mark=diamond*, dashed,mark size=1pt,line width=0.95pt, color=TolGreen] table[x=t, y=ABnq4] {\datatable};
    \addlegendentry{$AB,n_q=4$}
\end{axis}
\end{tikzpicture}

%% file: fig_numerical_phiTphi_inter.tex
\begin{tikzpicture}
\definecolor{TolBlue}{RGB}{0,114,178}
\definecolor{TolOrange}{RGB}{230,159,0}
\definecolor{TolGreen}{RGB}{0,158,115}
\begin{axis}[
  legend style={nodes={align=left}},
  legend cell align={left},
  font=\normalsize,
  width=\columnwidth, height=0.68\columnwidth, 
  xlabel={$t$}, ylabel={$\langle \Omega|\phi^\dagger(\bm 0)T_{0,\{0,1\}\to \{1,1\}}\phi(\bm 0)|\Omega\rangle$},
  ymin=-0.3, ymajorgrids,
  legend style={at={(0.02,0.98)},anchor=north west, draw=none, fill=none, font=\footnotesize},
  tick label style={font=\footnotesize},
  label style={font=\footnotesize},
  grid=both, grid style={line width=0.2pt, draw=black!15},
  clip=false
]

\addplot+[
  line width=0.95pt, mark=*, mark size=1.6pt, color=TolBlue
] table[
  row sep=\\, header=true, x=t, y=phidag_T_phi_VEV
]{
t phidag_T_phi_VEV\\
0 0\\
0.2 -0.01776\\
0.4 -0.02122\\
0.6 -0.00114\\
0.8 0.04309\\
1.0 0.10261\\
1.2 0.1612\\
1.4 0.19987\\
1.6 0.20251\\
1.8 0.16121\\
2 0.07933\\
};
\addlegendentry{Exact}

\addplot+[
  only marks, mark=square*, mark size=1.6pt,color=black!50,
  mark options={fill=black}
] table[
  row sep=\\, header=true, x=t, y=ReU
]{
t ReU \\
0 0.000974\\
0.2 -0.023565\\
0.4 -0.029099\\
0.6 -0.003149\\
0.8 0.05498\\
1 0.13291\\
1.2 0.2082\\
1.4 0.25489\\
1.6 0.25164\\
1.8 0.18904\\
2 0.073797\\
};
\addlegendentry{Adiabatic and Trotter, $n_q$=2}

\addplot+[
  only marks, mark=square*, mark size=1.6pt, color=TolOrange,
  error bars/.cd,
  y dir=both, y explicit
] table[
  row sep=\\, header=true, x=t, y=ReU, y error=Re_err
]{
t ReU Re_err\\
0 -0.11891 0.09375\\
0.2 0.04058 0.09795\\
0.4 0.11326 0.09224\\
0.6 -0.00494 0.08957\\
0.8 -0.00118 0.09124\\
1.0 -0.0208 0.08860\\
1.2 0.21748 0.09569\\
1.4 0.25375 0.09524\\
1.6 0.41377 0.09755\\
1.8 0.32375 0.09455\\
2.0 0.05140 0.09009\\
};
\addlegendentry{Simulation(Qiskit), $n_q$=2}

\end{axis}
\end{tikzpicture}